\documentstyle[multicol,epsf,aps,rotate]{revtex}

\begin{document}
\title{Systematic Computation of the Least Unstable
Periodic Orbits in Chaotic Attractors
}
\author{Fotis K. Diakonos$^{1,}$\footnote{E-mail: fdiakono@atlas.uoa.gr}, 
Peter Schmelcher$^{2,}$\footnote{E-mail: peter@tc.pci.uni-heidelberg.de} and        
Ofer Biham$^{3,}$\footnote{E-mail: biham@flounder.fiz.huji.ac.il}
}
\address{
$^{1}$
Department of Physics,
University of Athens,
GR-15771 Athens,
Greece
}
\address{
$^{2}$
Theoretische Chemie,      
Institut fuer Physikalische Chemie,  
Im Neuenheimer Feld 253,  
69120 Heidelberg,                    
Germany               
}
\address{
$^{3}$
Racah Institute of Physics, 
The Hebrew University, 
Jerusalem 91904, 
Israel
}
\maketitle

\begin{abstract}
We show that a recently proposed numerical technique for the 
calculation of unstable periodic orbits in chaotic attractors
is capable of finding the least unstable
periodic orbits of any given order. 
This is achieved by introducing a modified dynamical system
which has the same set of periodic orbits as the original chaotic system, 
but with a tuning parameter which is used to stabilize
the orbits selectively.
This technique is central 
for calculations using the stability criterion for the
truncation of cycle expansions, which 
provide highly improved convergence of calculations of
dynamical averages in generic chaotic attractors. 
The approach is demonstrated for the H\'enon attractor.
\end{abstract}


\begin{multicols}{2}

Unstable periodic orbits in chaotic attractors 
provide 
a useful hierarchical framework for calculations of
properties such as Lyapunov exponents, fractal dimensions
and entropies of the attractors
\cite{Grebogi88,Cvitanovic88a}. 
Periodic orbits have been used to characterize the attractors
in a variety of low dimensional dissipative dynamical systems,
including model systems such as discrete maps
\cite{Grebogi88,Cvitanovic88a}
as well as experimental time series
\cite{Schwartz92,Abarbanel93,Badii94,So96}.
In chaotic Hamiltonian systems, series expansions over unstable 
periodic orbits,
within the semi-classical approximation,
have been used to calculate the quantum energy level density
as well as properties of the wave functions
\cite{Gutzwiller90}.
Cycle expansion techniques
gave rise to highly improved convergence, particularly for systems in which
the symbolic dynamics is well understood and long periodic 
orbits are well shadowed by short ones
\cite{Cvitanovic88b,Artuso90}.

The calculation of unstable periodic orbits 
in chaotic dynamical systems
is a
difficult computational problem.
The difficulty is that in a chaotic system the numerical 
error grows exponentially with the length of the orbit.
Therefore, only short unstable periodic orbits can
be calculated using the standard techniques of map iteration.
Moreover, even if some orbits of a given order $p$ are calculated, one
has no guarantee that {\it all} orbits of this order have been found.
A numerical technique capable of calculating arbitrarily long
periodic orbits to any desired accuracy was introduced in Ref.
\cite{Biham89}
for the 
H\'enon map
\cite{Henon76}
and later applied to 
a variety of other dynamical systems
\cite{Biham90,Wenzel91,Biham92}.
Furthermore, this method provides a systematic framework for the
calculation of all periodic orbits of any given order $p$,
in which each orbit is identified by a unique binary symbol sequence 
$\{s_n\}$, $n=1,\dots,p$. 
The number of unstable periodic orbits of order $p$
for the H\'enon map
increases exponentially with
$p$ according to 
$N(p) = 2^{K_0 p}$,
where $K_0 \le 1$ is the topological entropy.
Therefore, the problem of finding {\it all} the periodic orbits
of order $p$ requires resources exponential in $p$.

Series expansions over periodic orbits used for calculations
of dynamical averages are typically ordered according to the orbit length
$p$
\cite{Grebogi88,Artuso90,Grassberger89,Biham92}.
Due to slow convergence 
and the fact that the number of orbits increases exponentially with $p$,
these series often require a huge number of orbits
\cite{Grassberger89,Biham92}.
It was recently proposed
\cite{Dettmann97a,Dettmann97b}
that using the cycle expansion framework for generic dynamical systems
one can obtain better convergence by truncating the expansion according
to the stability of the orbits 
\cite{Dahlqvist91}
rather than their length $p$.
This proposal is particularly useful since stability truncation does not
require detailed understanding of the symbolic dynamics, it tends to
preserve the shadowing properties and takes into account only the significant
orbits of each length
\cite{Dettmann97b}.

In this Letter we show that a
recently proposed technique  
\cite{Schmelcher97} 
provides a systematic framework for the calculation
of the least unstable periodic
orbits of any given order $p$. 
The resulting orbits are sorted according to their 
Lyapunov exponents starting with the least unstable.
The technique is highly flexible and can be applied in
a straightforward manner to a great variety of discrete
as well as continuous dynamical systems of any dimension.
Therefore, it opens the way for employing the proposal
of Refs.
\cite{Dettmann97a,Dettmann97b}
for a great variety of dynamical systems.

We will first describe the method.
Given a $N$-dimensional chaotic dynamical system
$U:~\vec{r}_{i+1}=\vec{f}(\vec{r}_i)$
we generate a set of 
dynamical
systems through the linear transformation:

\begin{equation}
S_k:~~\vec{r}_{i+1}=\vec{r}_i+\bbox{\Lambda}_k 
[\vec{f}(\vec{r}_i) - \vec{r}_i]
\label{dynsyss}
\end{equation}

\noindent
where
$\bbox{\Lambda}_k$ 
are invertible 
$N \times N$ 
constant matrices which can be cast in the form 
$\bbox{\Lambda}_k=\lambda \bbox{C}_k$ 
with 
$0 < \lambda < 1$. 
The matrices 
$\bbox{C}_k$ 
are orthogonal with
only one nonvanishing entry 
$(\pm 1)$ 
per row or column.
The systems 
$S_k$ 
are {\it equivalent} 
to the system $U$ in the sense that
there is a one to one correspondence between 
the fixed points of $U$ and those of 
$S_k$. 
The important advantage of the representation
of Eq. 
(\ref{dynsyss}) 
however is that, using a sufficiently small value for the
parameter 
$\lambda$, 
the fixed points of the transformed systems 
$S_k$ 
become stable and therefore can easily be determined 
by an iterative process. 
Furthermore, the radius of convergence of this 
iterative algorithm turns out to be finite. 
The above procedure can be easily extented to the higher iterates 
$\vec{f}^{(p)}(\vec{r})$ 
of 
$U$ 
[by replacing in Eq.
(\ref{dynsyss}) 
$\vec{f}$ 
with 
$\vec{f}^{(p)}$] 
allowing us to determine all the order $p$ cycles of $U$. 
The parameter 
$\lambda$ 
is a key quantity here.
For a given period
$p$, it operates 
as a filter allowing the selective stabilization of only those 
unstable periodic orbits, which possess
Lyapunov exponents smaller than a critical value.
Therefore, starting the search for unstable periodic orbits within a certain 
period
$p$
with a value of 
$\lambda \cong O(10^{-1})$ 
and gradually lowering 
$\lambda$ 
we obtain the list of all unstable orbits
of order $p$, 
starting with the least unstable orbit and
sorted with increasing values of their Lyapunov exponents.

Next, we
focus on
two dimensional systems. 
Denote the 
stability matrix 
of some periodic orbit of order $p$
by $M$, its matrix elements by
$m^{(p)}_{ij}$
where
$i,j=1,2$, 
and the 
stability  
eigenvalues of $M$
by
$\rho^{(p)}_{1}$ and $\rho^{(p)}_{2}$. 
Without loss of generality we assume 
$\vert \rho^{(p)}_1 \vert~ >~\vert \rho^{(p)}_2 \vert$.
In Refs.
\cite{Schmelcher97} 
a minimal set of matrices 
$\{\bbox{C}_k|~k=1,..,5~\}$,
which is 
necessary and sufficient to achieve stabilization for any 
kind of hyperbolic fixed points was provided.
The hyperbolic fixed points with reflection 
(namely, fixed points for which at least one of the eigenvalues  
$\rho^{(p)}_1$, 
$\rho^{(p)}_2$ 
is negative
while $|\rho^{(p)}_1| > 1$ and $|\rho^{(p)}_2| < 1$)
which satisfy $\rho^{(p)}_1 < 0$ 
become stable in the transformed system 
of Eq. 
(\ref{dynsyss}) 
if we use
$\bbox{\Lambda}_1=\lambda \bbox{C}_1$ 
with 
$\bbox{C}_1=\bbox{1}$,
where 
$\bbox{1}$
is the unit $2 \times 2$ matrix. 
After a little algebra a simple relation between the eigenvalues 
$\mu^{(p)}_{1,2}$ 
of the stabilized system $S_1$ and the eigenvalues 
$\rho^{(p)}_{1,2}$ 
in the original system $U$ can be obtained:

\begin{equation}
\mu^{(p)}_{1,2}= 1 - \lambda (1-\rho^{(p)}_{1,2}). 
\label{eigval}
\end{equation}

\noindent
The stability condition 
$\vert \mu^{(p)}_{1,2} \vert < 1$ 
leads to the critical value 
$\rho_c = 2 / \lambda - 1$
which represents an upper limit for the magnitude of the eigenvalues
$\rho^{(p)}_{1,2}$ 
of the fixed points
which are stabilized for a given 
$\lambda$. 
This means that only those orbits for which 
$\vert \rho^{(p)}_1 \vert < \rho_c$ 
become stable for the corresponding 
$\lambda$. 
Therefore,
varying $\lambda$ we can selectively 
extract those periodic orbits which 
possess stability eigenvalues less or 
equal to a given threshold value.

The stabilization process for the case of hyperbolic fixed points with 
reflection and $\rho^{(p)}_1 > 0$ or without reflection 
involves the matrices
$\{\bbox{C}_k~|~k=2,..,5\}$ \cite{Schmelcher97}. 
For these cases no exact monotonic
relationship like Eq. (\ref{eigval}) can be derived.
However, apart from exceptional cases (see below),
a selective stabilization procedure is possible similar to the case
of the matrix $\bbox{C}_1$: the overall tendency is again that
large values of $\lambda$ stabilize only the least unstable periodic
orbits and with decreasing value of $\lambda$ we get more and more
of the increasingly unstable periodic orbits.
To derive this let us consider without loss of generality the case 
$m^{(p)}_{11}~>~m^{(p)}_{22}$. 
The matrix which has to be used for the stabilization process
in this case is

\begin{equation} 
\bbox{C}_2=\left( \begin{array}{cc} -1 & 0 \\ 0 & 1 \end{array} \right).
\end{equation}

\noindent
The eigenvalues $\mu^{(p)}_{1,2}$ of the transformed system $S_2$
expressed in terms of the eigenvalues $\rho^{(p)}_{1,2}$ of the 
original system are

\begin{equation}
\mu^{(p)}_{1,2}= 1 - { \lambda \over 2}
\left[ {t \mp \sqrt{t^2 - 4 (\rho^{(p)}_1-1)(1 - \rho^{(p)}_2)}} \right]
\label{neweig}
\end{equation}

\noindent
where $t =  m_{11}^{(p)}-m_{22}^{(p)}$. 
Since typically
$\rho^{(p)}_1 \gg \rho^{(p)}_2$
one can use the approximation  
$Tr (M) = (m_{11}^{(p)}+m_{22}^{(p)}) = (\rho^{(p)}_1 + \rho^{(p)}_2)
\cong \rho^{(p)}_1$. 
Inserting this in the expressions for $\rho^{(p)}_{1,2}$ and using
$[\rho^{(p)}_{1}]^2 \gg 4 \rho^{(p)}_{1} \rho^{(p)}_{2}$
a detailed analysis of the possible cases leads to
the statement $\rho^{(p)}_{1} = \alpha m_{11}^{(p)}$ where $\alpha$
is a factor of order unity. 
As can be seen
from 
$\rho^{(p)}_{1,2}$ 
the only exception to this situation occurs if
${m_{22}^{(p)}}/{m_{11}^{(p)}} = -1 +\epsilon$, where $0 < \epsilon \ll 1$
which is certainly rare in chaotic systems.
We then have 
$t = \beta \rho_{1}^{(p)}$
with 
$\beta$ being a factor of $O(1)$. 
If the square root in 
Eq.(\ref{neweig}) 
is real we have 
$t^2 > 4 (\rho^{(p)}_1-1)(1 - \rho^{(p)}_2)$ 
and it immediately follows that the
relevant larger eigenvalue 
$\mu^{(p)}_1$ 
obeys 
$\mu^{(p)}_1 = 1 - {\lambda} \gamma \rho^{(p)}_1/2$ 
with $\gamma$ being a factor of 
$O(1)$.
If the square root is imaginary the real part of $\mu^{(p)}_1$,
which is responsible for the stabilization, obeys 
$Re(\mu^{(p)}_1)= 1 - {\lambda} \gamma \rho^{(p)}_1/2$. 
Exceptional cases
here are given by 
${m_{22}^{(p)}}/{m_{11}^{(p)}} = 1 - \epsilon$ where
$0 < \epsilon \ll 1$.
Although there is no strict monotonic ordering the above arguments
clearly demonstrate that a monotonic ordering occurs
to a good approximation. 
In addition we have performed a random matrix
simulation for the original stability matrices respecting
the above constraints due to hyperbolicity and 
calculated the distribution of the resulting prefactors $\gamma$
occurring in the eigenvalues of the transformed system.
The results of this analysis confirm the above-obtained
conclusions. The cases involving the other matrices
$\{\bbox{C}_k~\}$ can be treated analogously.

To demonstrate the power of the above method 
we now apply it to the H\'enon map.
To examine the spectrum of Lyapunov exponents we 
first calculated all the periodic orbits
up to $p=23$ 
for the H\'enon map
\cite{Henon76} 

\begin{equation}
x_{n+1}=a-x_n^2+b y_n, \ \ \ \ y_{n+1}=x_n
\label{henon}
\end{equation}

\noindent
with the parameters
$a=1.4$ and $b=0.3$,
using the method described in 
\cite{Biham89}.
The total number of orbits obtained is
118,407 (including cyclic permutations and repetitions of lower cycles). 
For each one of these orbits 
we calculated the Lyapunov exponent 
$h = \log(\vert \rho \vert)/p$
where $\rho$ 
(in absolute value)
is the largest eigenvalue
of the matrix 
$M = M_p \cdot \dots \cdot  M_2 \cdot  M_1$, 
where

\begin{equation}
M_n = \left( \begin{array}{cc}
         - 2 x_n & b\\
         1       & 0
         \end{array} \right)
\end{equation} 

\noindent
are the Jacobian matrices of the map.
The Lyapunov exponents 
of all orbits of order $p=1,\dots,23$
as a function of $p$ are shown in 
Fig. 1.
We observe that in this attractor the two fixed points have the largest
Lyapunov exponents and are thus the most unstable. 
The other Lyapunov
exponents form a band that becomes 
denser and 
broader as $p$ increases. We also observe a 
small number of orbits with unusually small Lyapunov exponents. 
Such orbits appear for orders 13, 16, 18 and 20.

To examine the dependence of Lyapunov exponents on the symbol 
sequence we plot the Lyapunov exponents for all the periodic 
orbits of order $p=20$
vs. the sequential number
of the orbit from 0 to $2^{20}-1$ 
(Fig. 2). 
The orbits are ordered such that for 
every two adjacent orbits in the plot the symbol sequences 
are different in only one bit. 
This allows us to examine how the
Lyapunov exponents change when one bit is switched in a long
symbol sequence. 
To achieve this the symbol sequence
$\{ s_n \}$
for each orbit is considered as a Gray code sequence
\cite{Kostopoulos75}.
A Gray to binary transformation 
$\{ s_n \} \rightarrow \{ s_n^{\prime} \}$
is then used and the decimal representation of
$\{ s_n^{\prime} \}$
is given in the horizontal axis of 
Fig. 2.

\begin{figure}
\narrowtext
\epsffile{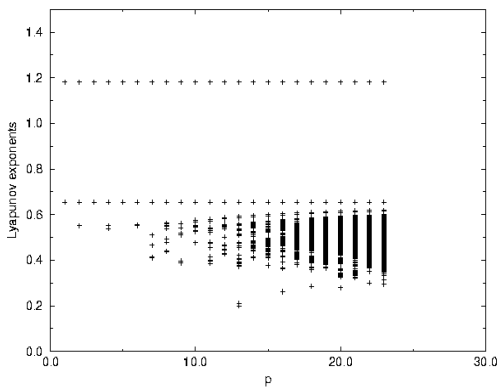}
\caption{
The Lyapunov exponents of all the periodic orbits of order 
$p=1,\dots,23$ 
as a function 
of $p$ for the H\'enon attractor at $a=1.4$ and  $b=0.3$. 
}
\end{figure}


\begin{figure}
\narrowtext
\epsffile{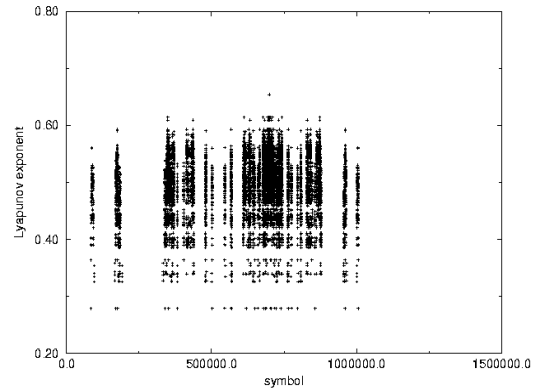}
\caption{
The Lyapunov exponents for all the orbits of order $p=20$ 
in the strange attractor at $a=1.4$ and $b=0.3$ as a function of 
the (decimal representation of the) 
symbol sequence, converted into the Gray code. 
The vacant domains
in the plot correspond to prunned periodic orbits.  
}
\end{figure}

To stabilize the periodic orbits in the
H\'enon attractor it turns out that
one needs
only two of the matrices 
$\bbox{C}_k$ 
namely 
$\bbox{C}_1=\bbox{1}$ 
and
$\bbox{C}_3 = - \bbox{C}_2$.
Using 
$\lambda$-values 
in the range 
$(0.05,0.002)$ 
and a set of $500$ starting
points on the attractor we were able to find, for each of the periods 
$p=1,\dots,23$, 
within a few seconds of 
computation on a desktop workstation,
the two least unstable periodic orbits. 
The results 
for $p=16,\dots,23$
are presented in 
Table 1 
which shows the period, 
the $(x,y)$ coordinates of one point of
each orbit and its 
Lyapunov exponent 
$h^{(p)}_1=({\ln \vert \rho^{(p)}_1 \vert})/{p}$.

\setcounter{table}{0}
\begin{narrowtext}
\begin{table}
\begin{tabular}{|c|c|c|c|} \hline \hline
Period & x-coord. & y-coord. &  Lowest Lyap. Exp. \\
\hline 
16 &  1.414441 &   0.525388  &  0.261873  \\ 
16 &  0.207904 &  -1.293373  &  0.259960  \\ 
17 &  1.168372 &   0.719909  &  0.380900  \\ 
17 &  0.956694 &   0.775439  &  0.379981  \\ 
18 & -1.255278 &   1.588681  &  0.285758  \\ 
18 & -1.256032 &   1.588953  &  0.284727  \\ 
19 &  0.475407 &   0.932648  &  0.365918  \\ 
19 &  0.608248 &   0.683230  &  0.365689  \\ 
20 &  0.999632 &   0.778785  &  0.279102  \\ 
20 &  0.687520 &  -0.496536  &  0.278732  \\ 
21 &  1.433765 &  -0.639919  &  0.323827  \\ 
21 &  1.018516 &   0.377002  &  0.323259  \\ 
22 &  1.184577 &   0.641833  &  0.300455  \\ 
22 &  0.641792 &   0.655127  &  0.300439  \\ 
23 &  1.416001 &   0.524053  &  0.295087  \\ 
23 &  1.596043 &  -0.481690  &  0.294950  \\ 
\hline \hline
\end{tabular}
\caption{The two periodic orbits with the lowest Lyapunov exponents for 
orders $p=16,\dots,23$, for the H\'enon map with
$a=1.4$ and $b=0.3$. The x and y coordinates of one point  
of each orbit are shown as well as 
the Lyapunov exponent of the orbit.
}
\end{table}
\end{narrowtext}

\begin{figure}
\narrowtext
\epsffile{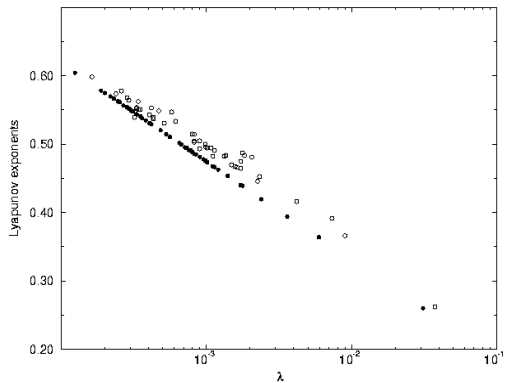}
\caption{The Lyapunov exponent of each periodic orbits of period 16 
for the H\'enon map $(a=1.4,b=0.3)$ is shown as a function 
of the critical value of the tuning parameter $\lambda$ below
which the orbit is stabilized.
Full circles represent orbits stabilized through $\bbox{C}_1$,
while open circles represent orbits stabilized through 
$\bbox{C}_3$.
For $\bbox{C}_1$ orbits, the Lyapunov exponents
are strictly monotonic vs. $\lambda$, while for 
$\bbox{C}_3$ orbits only the general trend is monotonic with some
deviations.
}
\end{figure}

To demonstrate how the periodic orbits are stabilized as the
tuning parameter 
$\lambda$ 
is decreased
we chose to present  
the results for
the period $p=16$. 
The total number of prime periodic
orbits 
(namely, not including cyclic permutations and repetitions of smaller cycles) 
for $p=16$ is 102. 
For a small value of  
$\lambda=10^{-4}$ 
and 500 starting points on the attractor we get all the 102 orbits stabilized.
To examine the stabilization process we start with
$\lambda=0.05$ and gradually lower it. 
This way, for each periodic orbit
we identify the critical value of 
$\lambda$ 
below which the orbit is stabilized. 
In Fig. 3, we present for all orbits of
period
$p=16$
the Lyapunov exponent of each orbit 
vs. the critical value of 
$\lambda$
below which this particular orbit is stabilized.
Orbits stabilized with the matrix 
$\bbox{C}_1$ 
are shown in full circles,
while orbits stabilized with
$\bbox{C}_3$ 
are shown in empty circles.
We observe that for orbits stabilized with
$\bbox{C}_1$
the Lyapunov exponents increase monotonically 
as $\lambda$ is lowered.
For orbits stabilized with
$\bbox{C}_3$
monotonicity is not strict, however the general trend is the same.
The nearly monotonic tendency of the Lyapunov exponents vs.
$\lambda$ demonstrates the suitability of our approach to determine,
with varying $\lambda$ the  set of least unstable periodic orbits
within a given period.

We observe that the spectrum of Lyapunov exponents 
(Fig. 2)
exhibits a rough
landscape, which resembles the energy spectrum 
which typically appears in hard minimization problems. 
Well known problems of this type are
finding spin glass ground states 
\cite{Mezard87}
and protein folding.
The use of combinatorial techniques for these problems is 
infeasible since the
computational resources required are exponential in the input size.
However, 
probabilistic algorithms such as simulated annealing
\cite{Kirkpatrick83}
can provide approximate results, which
for many applications are practically sufficient.
Our results may provide new insight about hard minimization problems in general.
The issue is whether one can artificially destabilize all the meta-stable states 
above some externally tuned energy $E$
and this way accelerate the convergence into the low-lying states.

In summary, we have shown that using the method proposed in Refs. 
\cite{Schmelcher97} 
one can obtain 
the periodic
orbits of any given order $p$ 
sorted according to their 
Lyapunov exponents starting with the least unstable one.
The method can be applied 
to a great variety of discrete
as well as continuous dynamical systems of any dimension.
Having the periodic orbits sorted in increasing order of their
Lyapunov exponents is highly useful in light of the 
recent proposal
\cite{Dettmann97a,Dettmann97b}
that in cycle expansion calculations for generic dynamical systems
better convergence can be obtained by truncating the expansion according
to the stability of the orbits rather than their length $p$.
In particular, stability truncation does not
require detailed understanding of the symbolic dynamics, it tends to
preserve the shadowing properties and takes into account only the significant
orbits of each length, leaving out an exponential number of insignificant 
orbits.

We thank P. Cvitanovi\'c for useful discussions.
F.K.D thanks the European Community for financial support.
P.S. acknowledges the hospitality of the Max-Planck Institute 
for Physics of Complex systems.




\end{multicols}

\end{document}